\documentclass[12pt]{article}

\usepackage{sbc-template}
\usepackage{graphicx,url}
\usepackage[T1]{fontenc}
\usepackage[utf8]{inputenc}
\usepackage{booktabs}
\usepackage{pgfplots}
\usepackage{multirow}
\usepackage{tikz}
\usepackage{amsmath}
\usepackage{float}
\usepackage{enumitem}
\usepackage{hyperref}

\pgfplotsset{compat=1.18}
\title{Simplifying Requirements Engineering in the Context of the LGPD: An LLM-Based Investigation}

\author{Cinara Gomes de Melo Carneiro, Renato de Freitas Bulcão Neto}

\address{Instituto de Informática -- Universidade Federal de Goiás (UFG)\\
  Caixa Postal 131 -- 74.690-900. - Goiânia -- GO -- Brazil \\
  \texttt{cinaragomes@discente.ufg.br, rbulcao@ufg.br}
}

\begin{document} 

\maketitle

\begin{abstract}
  Compliance with privacy legislation poses a complex challenge to Requirements Engineering (RE): translating legal norms into software requirements. In this context, this study investigates whether Large Language Models (LLMs) can simplify RE within the framework of the Brazilian General Data Protection Law (LGPD). The proposed approach utilizes current legislation to automatically generate User Stories and Acceptance Test Scenarios. The evaluation results demonstrated high performance, confirming the potential of LLMs to ensure regulatory compliance from the software's inception. 
\end{abstract}
      
\section{Introduction}

Compliance with data protection regulations, such as the General Data Protection Law (LGPD), has become a critical vector in software quality. In this context, Requirements Engineering (RE) assumes a central role in operationalizing Privacy by Design (PbD) \cite{cavoukian2009privacy}, a principle that requires the incorporation of privacy measures from system conception through to its implementation. Neglecting this initial stage can result in systems that violate fundamental rights, leading to severe legal sanctions and reputational damage \cite{LGPD}.

However, translating abstract legal norms into software requirements imposes significant challenges. Legal text is characterized by ambiguities, whereas software requires determinism. Exacerbating this scenario, software development professionals often lack the necessary legal knowledge to correctly interpret privacy laws, which hinders the proper implementation of protection mechanisms \cite{canedo2022privacy}\cite{alves2021especificaccao}. Although there are approaches based on Large Language Models (LLMs) for legal compliance in RE, a literature review reveals a significant gap in the Brazilian context. Almost all identified works focus exclusively on the European General Data Protection Regulation (GDPR). Furthermore, these solutions operate predominantly in a reactive manner, focusing on the \textit{a posteriori} verification of already consolidated artifacts, such as auditing privacy policies or analyzing source code \cite{hassani2024rethinking}\cite{garza2024privcomp}\cite{alecci2025toward}\cite{kunz2025using}, rather than assisting in their construction. Moreover, although the LGPD was inspired by European regulation, the direct importation of these approaches fails to capture local nuances and specific interpretations brought by the National Data Protection Authority (ANPD) in Brazil.

Given this scenario, this article presents an investigation into the potential of state-of-the-art LLMs to simplify RE in the privacy domain. Adopting an approach based on \textit{Design Science Research} \cite{hevner2004design}\cite{peffers2007design}, a Retrieval-Augmented Generation (RAG) based artifact was developed and evaluated. The system uses the text of the LGPD and ANPD guidelines as context to, through \textit{Few-Shot Prompting} and \textit{Chain-of-Thought} (CoT) techniques, assist engineers in the requirements elicitation, specification, and validation phases. The goal is to generate User Stories and Acceptance Test Scenarios that seek to ensure compliance from the beginning of software development, following the PbD principle.

This article is organized as follows: Section 2 analyzes works related to the intersection of RE, LLMs, and privacy legislation; Section 3 describes the theoretical foundation; Section 4 presents the Methodology, and Section 5 discusses the conclusion and future work.
 
\section{Related Work}

Recent literature indicates that the application of LLMs in Software Engineering has optimized regulatory compliance assurance. A large part of this scenario is dominated by \textit{a posteriori} verification strategies, where RAG techniques and Knowledge Graphs are employed to audit complex legal documents, such as contracts, Data Processing Agreements (DPAs), and privacy policies \cite{hassani2024rethinking, hassani2024enhancing, garza2024privcomp, jain2023automated}. In these ecosystems, approaches based on multi-agent architectures have been proposed to deal with the legal ambiguity of complex systems \cite{das2025multi}. Simultaneously, at the implementation level, approaches use LLMs to identify compliance with privacy legislation directly in technical artifacts, inspecting source code snippets according to official taxonomies \cite{kunz2025using} or analyzing low-level bytecode (\textit{Smali}) \cite{alecci2025toward}, always acting as a validation layer over already developed software. 

However, moving beyond mere document auditing, recent research has begun to explore active support for \textbf{requirements elicitation} through the direct generation of technical artifacts. Specifically, the method proposed by \cite{pragyan2024toward} operates in artifact generation by translating user narratives into structured segments of the Record of Processing Activities (RoPA) required by the GDPR. In the healthcare domain, the approach by \cite{madine2025leveraging} also actively acts in elicitation by synthesizing a list of atomic requirements and architectural constraints from the intersection between HIPAA and GDPR. Complementarily, when modeling new European directives (such as the \textit{AI Act}), \cite{falcarin2025legal} proposes generating Abstract Meaning Representations (AMR) and structured Knowledge Graphs to inform and guide the RE process. Despite these advances in artifact generation, these approaches fail to deliver operational requirements for agile development. Other methods operate under a purely extractive perspective of legal rights from predefined sources \cite{abualhaija2025llm} or are limited to evaluating ready-made user stories \cite{aberkane2024leveraging}, not advancing to the proactive synthesis of new functional requirements from a \textit{Privacy by Design} perspective.

The approach proposed in this article fills these gaps, standing out due to two fundamental differences from the state of the art. First, while most current works are limited to compliance verification, this method uses the capability of LLMs to translate legal obligations and transform them into fundamental artifacts of the agile development cycle: \textbf{User Stories} and \textbf{\textit{Gherkin}-based Acceptance Test Scenarios}. This anticipates regulatory compliance, actively integrating it into elicitation and delivering specifications ready for implementation. Second, while almost the entirety of the literature focused on privacy in RE and LLMs concentrates on the European GDPR, this work addresses the \textbf{Brazilian General Data Protection Law (LGPD)}, adapting requirements engineering to the legal and normative nuances of the Brazilian context.

\section{Theoretical Foundation}

\subsection{Method for Legal Requirements Extraction} 
\begin{figure}[H]
    \centering
    \includegraphics[width=0.9\linewidth]{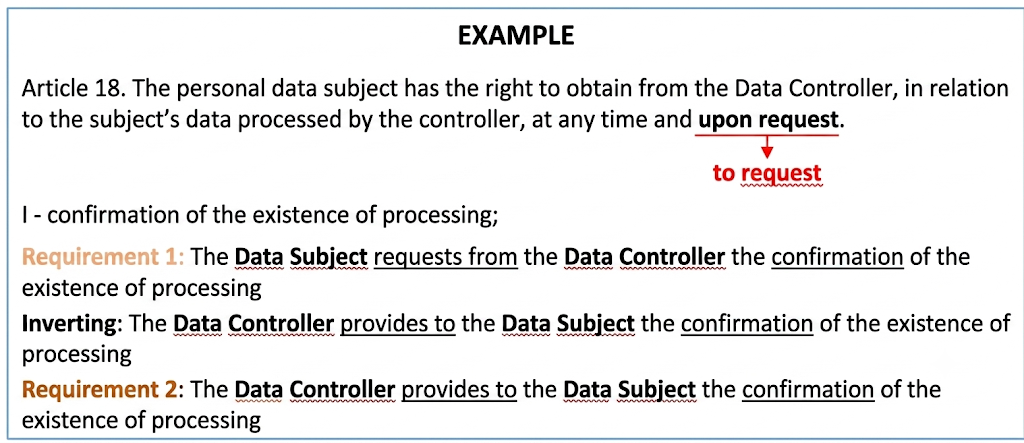}
    \caption{Method for Transforming Legal Requirements into Software Requirements}
    \label{fig:placeholder}
\end{figure}

The present work adopts the method proposed by \cite{de2024metodo}\cite{de2024evaluating}, which grounds the systematic translation of the LGPD's legal language into software requirements. The methodological core consists of the syntactic decomposition of legal provisions, building requirement sentences based on \textbf{Subject + Action Verb + Object}. The extraction process is segmented into the following phases: 

\begin{enumerate}[label=\textbf{\arabic*.}]
    \item \textbf{Actor Identification:} The initial step strictly locates the legal figures (Data Subject, Controller, Processor, and Data Protection Officer) defined in the norm. The actor appearing as the syntactic subject in the article's clause is designated as the actor in the requirement composition, avoiding the use of generic terms like "System".
    \item \textbf{Action Verb Location:} The verb defining the system's operation is identified. The method prescribes a critical normalization for implicit verbs: conditions described as nouns must be converted into action verbs. Example: The expression "upon request" is converted to the verb "request", clarifying the necessary action in the system.
    \item \textbf{Object Definition:} The term that suffers the verb's action is determined, consolidating the sentence in the form Actor + Action + Object.
    \item \textbf{Requirement Classification:} Items are categorized into Functional Requirements (FR), referring to behaviors and functionalities, or Non-Functional Requirements (NFR), referring to constraints and quality standards.
    \item \textbf{Refinement via Actor Inversion:} To maximize completeness, role inversion is applied. It verifies whether the actor opposite to the one identified in Step 1 should assume a complementary obligation. Example: A right attributed to the "Data Subject" generates, through inversion, an obligation to "allow" or "process" for the "Controller". Figure \ref{fig:placeholder} shows an output example generated by the method in question.
\end{enumerate}

In this work, the steps of the extraction method served as the basis for the chain of thought instructed to the model. By forcing the LLM to explicitly follow this extraction script — instead of directly requesting the final requirement — the system reproduces a human expert's analytical process, resulting in a requirement extraction that is more accurate and adherent to the legal text.

\subsection{Design Science Research}

The present research adopts Design Science Research (DSR) as its methodological approach, fundamentally oriented towards problem-solving and the creation of innovations that alter the state of the real world, differing from traditional sciences that only seek to describe or predict phenomena \cite{simon2019sciences}. The primary objective of this approach is the production of scientific knowledge through the development and evaluation of \textbf{artifacts}, which, according to \cite{hevner2004design}, are not limited to software, being classified into four categories: \textbf{constructs} (vocabulary), \textbf{models} (abstractions), \textbf{methods} (algorithms/practices), and \textbf{instantiations} (implemented systems). Conducting DSR requires a rigorous balance between the \textbf{Relevance} cycle, which aligns the research with business environment needs, and the \textbf{Rigor} cycle, which grounds the artifact in the existing theoretical knowledge base \cite{hevner2004design}. For operationalization, it follows the logical flow of the Design Science Research Methodology (DSRM) proposed by \cite{peffers2007design}, comprising the stages of \textbf{problem identification and motivation}, \textbf{definition of objectives}, \textbf{design and development}, \textbf{demonstration}, \textbf{evaluation}, and finally, \textbf{communication of results}, ensuring both the delivery of a practical solution and the formalization of academic knowledge.

\section{Methodology}

Regarding its nature, this investigation is classified as Applied Research, since it aims to generate knowledge for practical application directed at solving specific problems \cite{gil2008metodos}. The adopted approach is DSR, focused on the construction and evaluation of artifacts. The artifact developed in this work is an instantiation of the RAG architecture specialized in LGPD compliance.

The methodological process followed the steps proposed by \cite{peffers2007design}, as described below:

\subsection{Problem Identification and Motivation}

The first stage of this research consisted of identifying the difficulty professionals face in translating the LGPD's complex legal texts into software requirements. The gap motivating this work lies in the fact that current literature is predominantly focused on the European GDPR, with the vast majority of studies restricted to retrospective auditing, i.e., the validation of requirements in already existing systems. There is, therefore, a lack of approaches that enable the automatic generation of software requirements, ensuring legal fidelity from the software's conception.

\subsection{Definition of Solution Objectives}

The defined objective was to develop a system capable of receiving natural language queries and returning structured software requirements to investigate the potential of LLMs in RE activities. It was established that the artifact should generate output in the format of User Stories and Acceptance Test Scenarios according to the \textit{Gherkin} approach, ensuring coverage of the LGPD and ANPD norms.

\begin{figure}[H]
    \centering
    \includegraphics[width=1\linewidth]{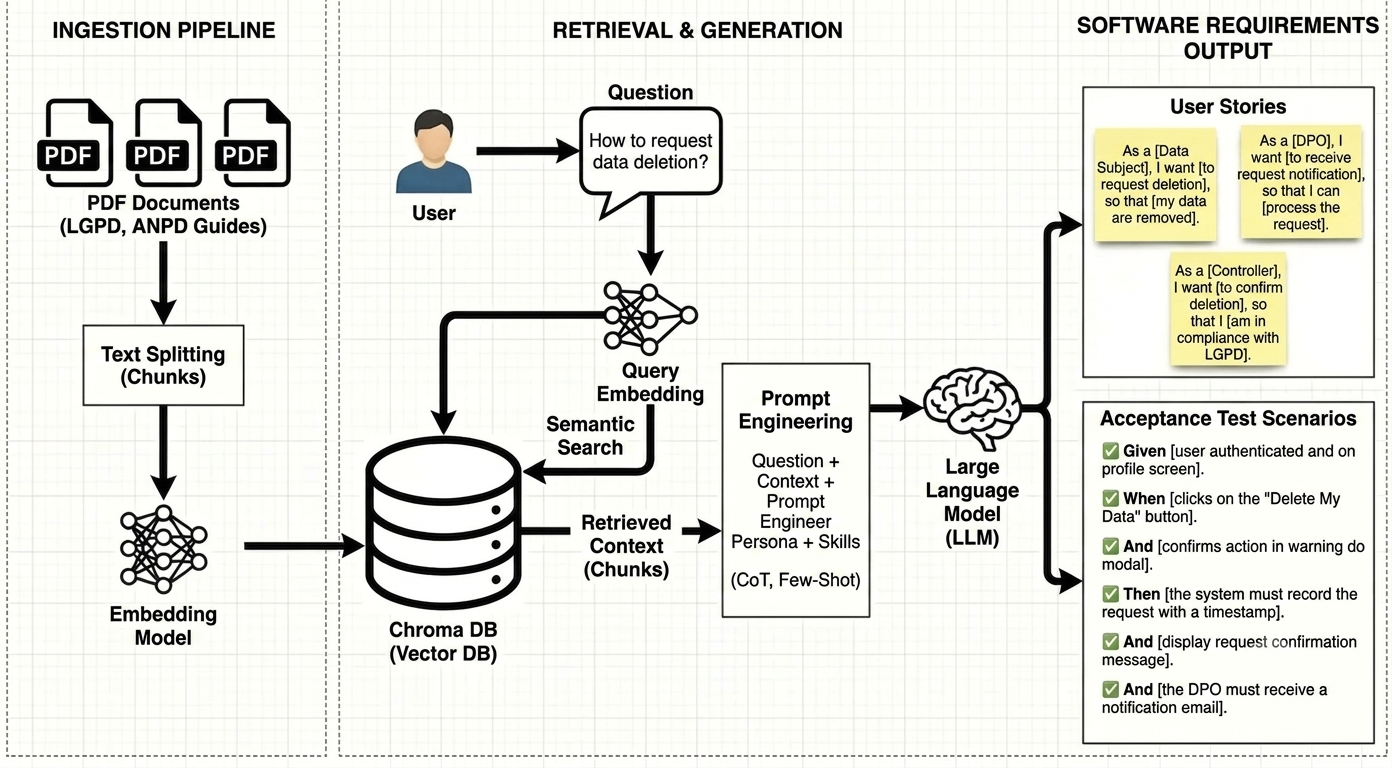}
    \caption{RAG Architecture in the context of the LGPD}
    \label{fig:RAGxLGPD}
\end{figure}

The diagram in Figure \ref{fig:RAGxLGPD} illustrates a RAG architecture designed to translate the LGPD's complexity into software requirements. The automated flow ingests raw legal documents, allows users to ask natural language questions about compliance procedures, and uses an LLM to generate software development artifacts. The final result is not just a legal answer, but User Stories and Acceptance Test Scenarios ready to be implemented by software developers so that systems are built in compliance from the start.

The construction of the artifact was carried out using the \textbf{Python} language, with data flow orchestration managed by the \textbf{LangChain} library. The technical architecture was implemented in the following layers:

\subsubsection{Data and Retrieval Layer}

The knowledge base was composed of the full text of the LGPD and ANPD guidelines in PDF format.

\begin{itemize}
    \item \textbf{Ingestion:} The \texttt{DirectoryLoader} integrated with \texttt{PyPDFLoader} was used for PDF extraction.
    \item \textbf{Advanced Chunking:} A \textit{Parent Document Retriever} architecture was applied, dividing documents into larger chunks (\textit{parents} with 1,500 characters) for context retention, and smaller chunks (\textit{children} with 400 characters) for vector search precision.
    \item \textbf{Embeddings and Vectorization:} To convert texts into vector representations, the multilingual model \texttt{intfloat/multilingual-e5-small} was used.
    \item \textbf{Storage:} Vectors were indexed in \textbf{Faiss}, and the original documents were kept in an \texttt{InMemoryStore}.
    \item \textbf{Re-ranking:} A refinement step was added using a \textit{Cross-Encoder} (\texttt{BAAI/bge-reranker-v2-m3}). The retrieved documents are re-evaluated and reordered to ensure maximum relevance to the legal context.
\end{itemize}

\subsubsection{Language Models}

The system was designed to be model-agnostic, simultaneously initializing multiple chains to allow direct comparison. For this study, the following were integrated:

\begin{enumerate}
    \item \textbf{Gemini Family (Google GenAI):} Using versions \textit{3.1 Flash Lite} and \textit{3.1 Pro Preview}.
    \item \textbf{Llama 3 Family (Hugging Face):} Using versions \textit{3.3 70B-Instruct} and \textit{3 8B-Instruct} executed via endpoints on the Hugging Face infrastructure.
\end{enumerate}

For execution, the following hyperparameters were defined:

\begin{itemize}
    \item \textbf{Temperature ($0.0$):} Selected with a null value to eliminate hallucination and ensure determinism, precision, and consistency in the answers, which is mandatory for the legal compliance context.
    \item \textbf{Search Top-K ($K=15 \rightarrow 5$):} Configured to initially retrieve 15 chunks, which are subsequently processed by the Re-ranker, strictly filtering the 5 most relevant documents (\textit{Final Top-K}) to compose the LLM's context window.
\end{itemize}

\subsubsection{Dual-Chain Architecture and Prompt Engineering}

Unlike traditional single-step approaches, orchestration was structured into a sequential chain pipeline (Dual-Chain) to mitigate confusion between legal interpretation and technical formatting:

\begin{enumerate}
    \item \textbf{Legal Analysis Chain (\textit{The Lawyer}):} A first extraction focused on describing the rights and rules of the law in plain text, using only the context retrieved by the RAG.
    \item \textbf{Requirements Engineering Chain (\textit{The Engineer}):} Receives the chewed legal text from the first chain and applies rigid constraints to convert it into a structured JSON of software requirements. 
\end{enumerate}

The effectiveness of this second stage relied on a \textit{System Prompt} combining two fundamental techniques recognized in the literature for improving reasoning and adherence in LLMs:

\begin{itemize}
    \item \textbf{Chain-of-Thought (CoT):} Studies demonstrate that the CoT technique significantly increases symbolic reasoning capacity \cite{santos2024requirements}. The model is instructed to report its thought process (\texttt{\_thought\_process}) validating the coverage of retrieved clauses. The prompt instructs strict semantic decomposition into the triad: \textit{Actor (Controller, Data Subject, Processor, Data Protection Officer)} + \textit{Action Verb} + \textit{Object} \cite{de2024metodo}, categorically prohibiting the use of generic actors like "System".
    
    \item \textbf{Few-Shot Learning:} Relying on \cite{pragyan2024toward}, which evidenced that providing examples is crucial for correct requirements extraction. A validated input-output JSON (\textit{few-shot}) was incorporated into the prompt, ensuring syntactic standardization of keys, User Stories, and Scenarios in \textit{Gherkin} format, further ensuring safe output parsing through the \texttt{json\_repair} library.
\end{itemize}

\subsection{Experimental Evaluation}

The evaluation stage imposes challenges that transcend simple textual similarity verification. As discussed by \cite{yu2024evaluation}, RAG systems should not be treated as "black boxes", since inconsistencies in the final response can derive from failures in context retrieval, model reasoning, or the generation stage. 

To evaluate the generated artifacts — which encompass User Stories, Gherkin Test Scenarios, and LGPD legal references —, an approach based on the \textit{LLM-as-a-judge} paradigm was chosen, ensuring scalability and consistency in the validation cycle. In this paradigm, an advanced model acts independently to evaluate outputs under the critical pillars of Retrieval and Generation, using the \textbf{RAGAS (Retrieval Augmented Generation Assessment)} framework \cite{es2024ragas}.

To act as the algorithmic judge in this evaluation, the Llama 3.3 70B model was employed. The adoption of this model is based on the recommendations of \cite{pradhan2025ragevalx}, which indicate its effectiveness specifically for answer relevance and context relevance metrics. However, aiming to simplify and standardize the evaluation pipeline, Llama 3.3 70B was expanded and used to calculate all metrics in this study. Furthermore, to ensure the accuracy and reliability of the mathematical measurement, the ground truth used as a reference by the judge LLM was prepared and rigorously validated by two LGPD specialists.

\subsubsection{Test Cases}

The system was subjected to a comparative evaluation using five reference queries based on questions of high relevance for LGPD compliance, aiming at the extraction of functional and non-functional requirements:

\begin{enumerate}
    \item \textbf{Q1:} ``What are the technical security measures for access control?''
    \item \textbf{Q2:} ``For what purposes is the retention of personal data authorized?''
    \item \textbf{Q3:} ``In which hypotheses will the termination of personal data processing occur?''
    \item \textbf{Q4:} ``In which hypotheses may the processing of sensitive data occur?''
    \item \textbf{Q5:} ``List all data subject rights described in Art. 18 of the LGPD.''
\end{enumerate}

Each query resulted in the generation of multiple software requirements, varying in quantity according to the tested model. The repository with the generated outputs is available at \url{https://huggingface.co/datasets/CinaraMelo/resultado_avaliacao_ragas/blob/main/respostas_modelos.md}

\subsubsection{Protocol and Metrics Definition}

Based on the taxonomy presented by \cite{es2024ragas}, metrics were automated to independently quantify the performance of RAG systems. The judge LLM receives the RAG answers and the retrieved contexts, applying the following fundamental criteria:

\paragraph{Generation Metrics}
These metrics evaluate the generation module's ability to process the provided context and structure a high-quality response \cite{es2024ragas}.

\begin{itemize}
    \item \textbf{Faithfulness:} Evaluates whether the generated requirement is factually grounded in the context retrieved from the LGPD, ensuring the absence of hallucinations. The score $F$ is computed as \cite{es2024ragas}:
    
    \begin{equation}
        F = \frac{|V|}{|S|}
    \end{equation}
    
    \noindent Where $|S|$ is the total number of statements extracted from the answer and $|V|$ is the number of statements confirmed by the LLM as supported by the context.

    \item \textbf{Answer Relevance:} Assesses whether the produced technical requirement directly addresses the formulated question, penalizing redundant or incomplete outputs \cite{es2024ragas}. The $AR$ metric is defined as:
    
    \begin{equation}
        AR = \frac{1}{n} \sum_{i=1}^{n} \text{sim}(q, q_i)
    \end{equation}
    
    \noindent Where $\text{sim}(q, q_i)$ represents the cosine similarity between the embeddings of the original question $q$ and the $n$ hypothetical questions $q_i$ generated from the answer.
\end{itemize}

\paragraph{Retrieval Metrics}
These metrics evaluate the system's effectiveness in identifying and ordering pertinent context passages in the vector base.

\begin{itemize}
    \item \textbf{Context Precision:} Evaluates the quality of the retrieval results' ordering, verifying whether the relevant elements (\textit{ground truth}) are ranked in the top positions. It is calculated as:
    
    \begin{equation}
        \text{Context Precision} = \frac{\sum_{k=1}^{n} (\text{Precision}@k \times v_k)}{\sum_{k=1}^{n} v_k}
    \end{equation}
    
    \noindent Where $k$ represents the position in the list, $n$ the total number of retrieved items, $v_k$ is a relevance indicator (1 if the item at position $k$ is relevant, 0 otherwise), and $\text{Precision}@k$ is the precision up to position $k$.

    \item \textbf{Context Recall:} Verifies whether the system retrieved all the essential chunks to compose a normatively complete answer, comparing the retrieved context with the \textit{Ground Truth} of each question \cite{es2024ragas}.
\end{itemize}

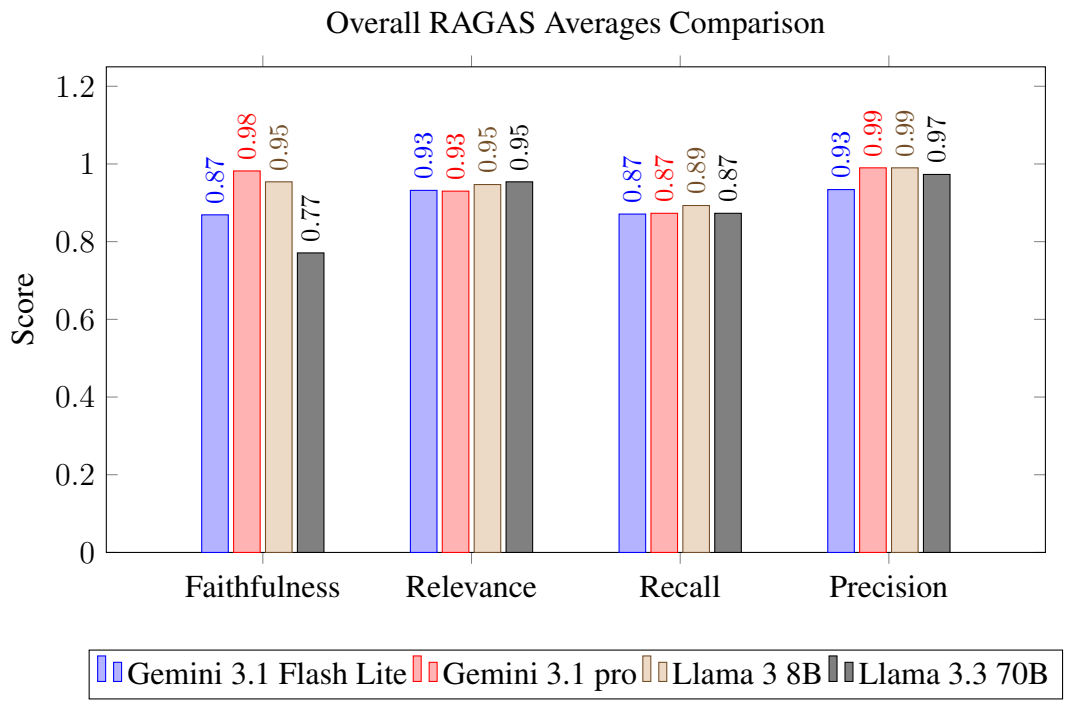
\begin{figure}[htbp]
\centering
\begin{tikzpicture}
\begin{axis}[
    ybar,
    enlarge x limits=0.25,
    legend style={at={(0.5,-0.2)},
      anchor=north,legend columns=-1},
    ylabel={Score},
    symbolic x coords={Faithfulness,Relevance,Recall,Precision},
    xtick=data,
    nodes near coords,
    nodes near coords style={rotate=90, anchor=west, font=\footnotesize},
    ymin=0, ymax=1.25, 
    width=14cm,        
    height=8cm,
    bar width=10pt,    
    title={Overall RAGAS Averages Comparison}
]
    \addplot coordinates {(Faithfulness,0.869) (Relevance,0.932) (Recall,0.871) (Precision,0.934)};
    \addlegendentry{Gemini 3.1 Flash Lite}
    \addplot coordinates {(Faithfulness,0.982) (Relevance,0.930) (Recall,0.873) (Precision,0.990)};
    \addlegendentry{Gemini 3.1 pro}
    \addplot coordinates {(Faithfulness,0.954) (Relevance,0.947) (Recall,0.893) (Precision,0.990)};
    \addlegendentry{Llama 3 8B}
    \addplot coordinates {(Faithfulness,0.771) (Relevance,0.954) (Recall,0.873) (Precision,0.973)};
    \addlegendentry{Llama 3.3 70B}
\end{axis}
\end{tikzpicture}
\caption{Comparison of overall averages obtained for each RAGAS metric.}
\label{fig:medias_ragas}
\end{figure}

Table \ref{tab:detalhes_questoes} details the performance of the Gemini 3.1 Flash Lite, Gemini 3.1 Pro, Llama 3.1 8B, and Llama 3.3 70B models regarding the aforementioned metrics for each query. Additionally, graph \ref{fig:medias_ragas} illustrates the overall average performance among the compared models.

\begin{table}[htbp]
\centering
\caption{Detailed results of each question by model.}
\label{tab:detalhes_questoes}
\begin{tabular}{llcccc}
\toprule
\textbf{Model} & \textbf{Question} & \textbf{Faithfulness} & \textbf{Relevance} & \textbf{Recall} & \textbf{Precision} \\
\midrule
    Gemini 3.1 Flash Lite & Q1 & 0.8000 & 0.9260 & 0.6667 & 0.8056 \\
     & Q2 & 1.0000 & 0.8968 & 1.0000 & 1.0000 \\
     & Q3 & 0.8750 & 0.9669 & 0.8000 & 1.0000 \\
     & Q4 & 0.9474 & 0.9389 & 0.8889 & 0.9500 \\
     & Q5 & 0.7222 & 0.9330 & 1.0000 & 0.9167 \\
    \midrule
    Gemini 3.1 pro & Q1 & 1.0000 & 0.9534 & 0.6667 & 1.0000 \\
     & Q2 & 0.9565 & 0.9316 & 0.8000 & 1.0000 \\
     & Q3 & 1.0000 & 0.9192 & 1.0000 & 1.0000 \\
     & Q4 & 1.0000 & 0.9435 & 0.9000 & 0.9500 \\
     & Q5 & 0.9524 & 0.9018 & 1.0000 & 1.0000 \\
    \midrule
    Llama 3 8B & Q1 & 0.8333 & 0.9417 & 0.6667 & 1.0000 \\
     & Q2 & 1.0000 & 0.9485 & 0.8000 & 1.0000 \\
     & Q3 & 0.9375 & 0.9774 & 1.0000 & 1.0000 \\
     & Q4 & 1.0000 & 0.9513 & 1.0000 & 0.9500 \\
     & Q5 & 1.0000 & 0.9177 & 1.0000 & 1.0000 \\
    \midrule
    Llama 3.3 70B & Q1 & 0.7037 & 1.0000 & 0.6667 & 1.0000 \\
     & Q2 & 1.0000 & 0.9526 & 0.8000 & 1.0000 \\
     & Q3 & 0.9524 & 0.9380 & 1.0000 & 1.0000 \\
     & Q4 & 0.9286 & 0.9648 & 0.9000 & 0.9500 \\
     & Q5 & 0.2727 & 0.9133 & 1.0000 & 0.9167 \\
\bottomrule
\end{tabular}
\end{table}

\section{Conclusion}

The analysis validates the integration of LLMs with the RAG architecture as an effective tool for Requirements Engineering, presenting robust retrieval and generation metrics that preserved content accuracy and the legal validity of information. Although the four evaluated models reduced cognitive load, Gemini 3.1 Pro and Llama 3 8B stood out by ensuring the highest Precision and Faithfulness to the context. It should be noted that, in cases where responses showed a lower score in the faithfulness metric, it was observed that the LLM generated a volume of information superior to that contained in the reference ground truth. This level of detail led the evaluator judge to interpret the excess content as hallucination. However, an analysis revealed that much of this additional information was, in fact, correct and pertinent, being strictly penalized for extrapolating the restricted scope of the ground truth. It is also concluded that the performance of proprietary models was statistically similar to that of open-source alternatives. This demonstrates that it is possible to achieve generation and context retrieval quality using \textit{open-source} models, which represents a significant advantage in terms of technical feasibility, cost reduction, and data privacy.

This study investigated the potential of LLMs in simplifying the extraction and adaptation of privacy requirements, aiming to support RE. The comparative tests conducted between \textit{open-source} and proprietary models indicated that adopting the RAG architecture constitutes a promising approach for this purpose. Generally, it was verified that the performance of the evaluated models showed similarity, reinforcing the technical feasibility of using open solutions in real scenarios.

Corroborating the findings of \cite{das2025multi}, it was observed that applying a \textit{re-ranking} mechanism was a determining factor in maintaining high levels of relevance and faithfulness in the retrieved context, although its use imposes a high computational cost on the process. It must be noted, however, that the stochastic nature of language generation by these models implies an inherent randomness, which can cause slight fluctuations, up or down, in evaluation metrics during different executions. Moreover, experimentation evidenced that the quality of \textit{prompt} engineering directly and significantly interferes with the accuracy of the obtained responses. Despite the promising performance, it was noted as a limitation that the LLMs presented difficulties and made errors when describing and correctly referencing citations from legal texts.

Finally, as a direction for future work, we intend to expand the scope of this research to evaluate the intersection of requirements between the LGPD and the Statute of the Child and Adolescent in the digital environment (ECA Digital). To mitigate the instructional limitations found, we also plan to improve \textit{prompt} engineering through the adoption of optimization frameworks, such as DSPy, aiming to evaluate whether this approach provides a measurable improvement in response quality and citation accuracy. Furthermore, following the methodological guidelines proposed by Pradhan \textit{et al.} \cite{pradhan2025ragevalx}, we plan to adopt multiple LLMs acting as judges (\textit{LLM-as-a-Judge}), aiming to increase the robustness of quantitative validation. In addition to this automated validation, qualitative evaluations with human domain experts — covering both Requirements Engineering and data protection professionals — are planned to attest to the practical applicability, completeness, and strict compliance of the generated artifacts.

\bibliographystyle{sbc}

\begin{thebibliography}{}

\bibitem[Aberkane et~al. 2024]{aberkane2024leveraging}
Aberkane, A.-J., vanden Broucke, S., Poels, G., and Georgiadis, G. (2024).
\newblock Leveraging chatgpt for gdpr compliance in requirements engineering: A pilot study.
\newblock In {\em 2024 IEEE International Conference on Security, Privacy, Anonymity in Computation and Communication and Storage (SpaCCS)}, pages 34--41. IEEE.

\bibitem[Abualhaija et~al. 2025]{abualhaija2025llm}
Abualhaija, S., Ceci, M., Sannier, N., Bianculli, D., Lannier, S., Siclari, M., Voordeckers, O., and Tosza, S. (2025).
\newblock Llm-assisted extraction of regulatory requirements: A case study on the gdpr.
\newblock In {\em 2025 IEEE 33rd International Requirements Engineering Conference (RE)}, pages 142--154. IEEE.

\bibitem[Alecci et~al. 2025]{alecci2025toward}
Alecci, M., Sannier, N., Ceci, M., Abualhaija, S., Samhi, J., Bianculli, D., Bissyand{\'e}, T., and Klein, J. (2025).
\newblock Toward llm-driven gdpr compliance checking for android apps.
\newblock In {\em Proceedings of the 33rd ACM International Conference on the Foundations of Software Engineering}, pages 606--610.

\bibitem[Alves and Neves 2021]{alves2021especificaccao}
Alves, C. and Neves, M. (2021).
\newblock Especifica{\c{c}}{\~a}o de requisitos de privacidade em conformidade com a {LGPD}: Resultados de um estudo de caso.
\newblock In {\em WER}.

\bibitem[Brasil 2018]{LGPD}
Brasil (2018).
\newblock Lei geral de proteção de dados pessoais.
\newblock \url{https://www.planalto.gov.br/ccivil_03/_ato2015-2018/2018/lei/l13709.htm}.
\newblock Acesso em: 25 de outubro de 2025.

\bibitem[Canedo et~al. 2022]{canedo2022privacy}
Canedo, E.~D., Bandeira, I.~N., Calazans, A. T.~S., Costa, P. H.~T., Can{\c{c}}ado, E. C.~R., and Bonif{\'a}cio, R. (2022).
\newblock Privacy requirements elicitation: a systematic literature review and perception analysis of it practitioners.
\newblock {\em Requirements Engineering}, pages 1--18.

\bibitem[Cavoukian et~al. 2009]{cavoukian2009privacy}
Cavoukian, A. et~al. (2009).
\newblock Privacy by design: The 7 foundational principles.
\newblock {\em Information and privacy commissioner of Ontario, Canada}, 5:12.

\bibitem[Das et~al. 2025]{das2025multi}
Das, S., Deb, N., Chaki, N., and Cortesi, A. (2025).
\newblock A multi-agent rag framework for regulatory compliance checking of software requirements.
\newblock {\em ACM Transactions on Software Engineering and Methodology}.

\bibitem[de~Melo~Carneiro et~al. 2024a]{de2024evaluating}
de~Melo~Carneiro, C.~G., Kudo, T.~N., and Bulc{\~a}o-Neto, R.~F. (2024a).
\newblock Evaluating privacy requirement patterns based on the brazilian general personal data protection law.
\newblock In {\em Proceedings of the XXIII Brazilian Symposium on Software Quality}, pages 114--124.

\bibitem[de~Melo~Carneiro et~al. 2024b]{de2024metodo}
de~Melo~Carneiro, C.~G., Kudo, T.~N., and Neto, R. F.~B. (2024b).
\newblock Um m{\'e}todo para transforma{\c{c}}{\~a}o de requisitos legais em padr{\~o}es de requisitos de software: Um estudo com a lgpd.
\newblock In {\em Congresso Ibero-Americano em Engenharia de Software (CIbSE)}, pages 348--355. SBC.

\bibitem[Es et~al. 2024]{es2024ragas}
Es, S., James, J., Anke, L.~E., and Schockaert, S. (2024).
\newblock Ragas: Automated evaluation of retrieval augmented generation.
\newblock In {\em Proceedings of the 18th Conference of the European Chapter of the Association for Computational Linguistics: System Demonstrations}, pages 150--158.

\bibitem[Falcarin et~al. 2025]{falcarin2025legal}
Falcarin, P., Chowdhury, P., Carbone, E., Scantamburlo, T., Tripodi, R., and Vascon, S. (2025).
\newblock Legal requirements compliance using nlp and knowledge graphs.
\newblock In {\em 2025 IEEE 33rd International Requirements Engineering Conference Workshops (REW)}, pages 412--418. IEEE.

\bibitem[Garza et~al. 2024]{garza2024privcomp}
Garza, L., Elluri, L., Piplai, A., Kotal, A., Gupta, D., and Joshi, A. (2024).
\newblock Privcomp-kg: Leveraging kg and llm for compliance verification.
\newblock In {\em 2024 IEEE 6th International Conference on Trust, Privacy and Security in Intelligent Systems, and Applications (TPS-ISA)}, pages 97--106. IEEE.

\bibitem[Gil 2008]{gil2008metodos}
Gil, A.~C. (2008).
\newblock {\em M{\'e}todos e t{\'e}cnicas de pesquisa social}.
\newblock 6. ed. Ediitora Atlas SA.

\bibitem[Hassani 2024]{hassani2024enhancing}
Hassani, S. (2024).
\newblock Enhancing legal compliance and regulation analysis with large language models.
\newblock In {\em 2024 IEEE 32nd International Requirements Engineering Conference (RE)}, pages 507--511. IEEE.

\bibitem[Hassani et~al. 2024]{hassani2024rethinking}
Hassani, S., Sabetzadeh, M., Amyot, D., and Liao, J. (2024).
\newblock Rethinking legal compliance automation: Opportunities with large language models.
\newblock In {\em 2024 IEEE 32nd International Requirements Engineering Conference (RE)}, pages 432--440. IEEE.

\bibitem[Hevner et~al. 2004]{hevner2004design}
Hevner, A.~R., March, S.~T., Park, J., and Ram, S. (2004).
\newblock Design science in information systems research.
\newblock {\em MIS quarterly}, pages 75--105.

\bibitem[Jain et~al. 2023]{jain2023automated}
Jain, C., Anish, P.~R., and Ghaisas, S. (2023).
\newblock Automated identification of security and privacy requirements from software engineering contracts.
\newblock In {\em 2023 IEEE 31st International Requirements Engineering Conference Workshops (REW)}, pages 234--238. IEEE.

\bibitem[Kunz et~al. 2025]{kunz2025using}
Kunz, I., Kao, C.-Y., Kowatsch, D., Hiller, J., Sch{\"u}tte, J., Prokhorenkov, D., and Bettinger, K. (2025).
\newblock Using llms to identify personal data processing in source code.
\newblock In {\em 2025 IEEE Security and Privacy Workshops (SPW)}, pages 137--144. IEEE.

\bibitem[Madine et~al. 2025]{madine2025leveraging}
Madine, M., Musamih, A., Salah, K., Zemerly, M., Ellahham, S., and Jayaraman, R. (2025).
\newblock Leveraging llms for rapid development of regulatory-compliant software in healthcare.
\newblock In {\em 2025 IEEE Technology and Engineering Management Society Conference-Global (TEMSCON Global)}, pages 1--6. IEEE.

\bibitem[Peffers et~al. 2007]{peffers2007design}
Peffers, K., Tuunanen, T., Rothenberger, M.~A., and Chatterjee, S. (2007).
\newblock A design science research methodology for information systems research.
\newblock {\em Journal of management information systems}, 24(3):45--77.

\bibitem[Pradhan 2025]{pradhan2025ragevalx}
Pradhan, R. (2025).
\newblock Ragevalx: An extended framework for measuring core accuracy, context integrity, robustness, and practical statistics in rag pipelines.
\newblock {\em International Journal of Computer Technology and Electronics Communication}, 8(5):11305--11311.

\bibitem[Pragyan et~al. 2024]{pragyan2024toward}
Pragyan, K., Ghandiparsi, R., Slavin, R., Ghanavati, S., Breaux, T., and Hosseini, M.~B. (2024).
\newblock Toward regulatory compliance: A few-shot learning approach to extract processing activities.
\newblock In {\em 2024 IEEE 32nd International Requirements Engineering Conference Workshops (REW)}, pages 241--250. IEEE.

\bibitem[Santos et~al. 2024]{santos2024requirements}
Santos, S., Breaux, T., Norton, T., Haghighi, S., and Ghanavati, S. (2024).
\newblock Requirements satisfiability with in-context learning.
\newblock In {\em 2024 IEEE 32nd International Requirements Engineering Conference (RE)}, pages 168--179. IEEE.

\bibitem[Simon 2019]{simon2019sciences}
Simon, H.~A. (2019).
\newblock {\em The Sciences of the Artificial, reissue of the third edition with a new introduction by John Laird}.
\newblock MIT press.

\bibitem[Yu et~al. 2024]{yu2024evaluation}
Yu, H., Gan, A., Zhang, K., Tong, S., Liu, Q., and Liu, Z. (2024).
\newblock Evaluation of retrieval-augmented generation: A survey.
\newblock In {\em CCF Conference on Big Data}, pages 102--120. Springer.

\end{thebibliography}

\end{document}